\documentclass[prb,twocolumn,twoside,superscriptaddress,showkeys,preprintnumbers,amsmath,amssymbaps,10pt,notitlepage,showpacs,superscriptaddress,floatfix]{revtex4-1}

\usepackage{lipsum}
\makeatletter
\newcommand*{\balancecolsandclearpage}{%
	\close@column@grid
	\clearpage
	\begin{center}
		\textbf{\Large Supplemental Materials}
	\end{center}
	\vspace{5 mm}
	Here we summarize details on the theoretical modeling of electrostatic effects and inter-layer tunneling in graphene-LaAlO$_3$/SrTiO$_3$ hybrid systems.
	\vspace{15 mm}
	\twocolumngrid}

\makeatother

\usepackage{lmodern}
\usepackage{bm}
\usepackage{color}
\usepackage{pdfsync}
\usepackage[percent]{overpic}
\usepackage{graphicx}
\usepackage{dcolumn}
\usepackage{bm}
\usepackage{amsmath}
\usepackage{amssymb}
\usepackage{natbib}
\usepackage[normalem]{ulem}
\usepackage[breaklinks=false,colorlinks,citecolor=blue,linkcolor=blue,urlcolor=blue]{hyperref}
\usepackage[usenames,dvipsnames]{xcolor}

\begin{document}

\title{Transport in strongly-coupled graphene-LaAlO$_3$/SrTiO$_3$ hybrid systems}

\author{I.~Aliaj}
\email{ilirjan.aliaj@sns.it}
\affiliation{NEST, Scuola Normale Superiore, I-56126 Pisa,~Italy}

\author{I.~Torre}
\affiliation{NEST, Scuola Normale Superiore, I-56126 Pisa,~Italy}

\author{V.~Miseikis}
\affiliation{Center for Nanotechnology Innovation @NEST, Istituto Italiano di Tecnologia, Piazza San Silvestro 12, 56127 Pisa,~Italy}

\author{E.~di~Gennaro}
\affiliation{CNR-SPIN and Dipartimento di Fisica, Complesso Universitario di Monte S.Angelo, Via Cintia, 80126 Naples,~Italy}

\author{A.~Sambri}
\affiliation{CNR-SPIN and Dipartimento di Fisica, Complesso Universitario di Monte S.Angelo, Via Cintia, 80126 Naples,~Italy}

\author{A.~Gamucci}
\affiliation{Istituto Italiano di Tecnologia, Graphene Labs, Via Morego 30, I-16163 Genova,~Italy}

\author{C.~Coletti}
\affiliation{Center for Nanotechnology Innovation @NEST, Istituto Italiano di Tecnologia, Piazza San Silvestro 12, 56127 Pisa,~Italy}
\affiliation{Istituto Italiano di Tecnologia, Graphene Labs, Via Morego 30, I-16163 Genova,~Italy}

\author{F.~Beltram}
\affiliation{NEST, Istituto Nanoscienze-CNR and Scuola Normale Superiore, I-56126 Pisa,~Italy}

\author{F.~M.~Granozio}
\affiliation{CNR-SPIN and Dipartimento di Fisica, Complesso Universitario di Monte S.Angelo, Via Cintia, 80126 Naples,~Italy}

\author{M.~Polini}
\affiliation{Istituto Italiano di Tecnologia, Graphene Labs, Via Morego 30, I-16163 Genova,~Italy}
\affiliation{NEST, Scuola Normale Superiore, I-56126 Pisa,~Italy}

\author{V.~Pellegrini}
\affiliation{Istituto Italiano di Tecnologia, Graphene Labs, Via Morego 30, I-16163 Genova,~Italy}

\author{S.~Roddaro}
\email{stefano.roddaro@sns.it}
\affiliation{NEST, Istituto Nanoscienze-CNR and Scuola Normale Superiore, I-56126 Pisa,~Italy}

\nocite{*}
\begin{abstract}
We report on the transport properties of hybrid devices obtained by depositing graphene on a LaAlO$_3$/SrTiO$_3$ oxide junction hosting a $4~{\rm nm}$-deep two-dimensional electron system. At low graphene-oxide inter-layer bias the two electron systems are electrically isolated, despite their small spatial separation, and very efficient reciprocal gating is shown. A pronounced rectifying behavior is observed for larger bias values and ascribed to the interplay between electrostatic depletion and tunneling across the LaAlO$_3$ barrier. The relevance of these results in the context of strongly-coupled bilayer systems is discussed.
\end{abstract}


\maketitle
	
Graphene is widely investigated in view of possible device applications owing to its excellent, electric-field tunable electronic properties~\cite{rev-gr}, its chemical and structural robustness, its ease of production and integrability with a plethora of other material systems~\cite{rev-gr-proc}. In addition, because of its single-atom thickness, its properties can be very sensitive to the local environment. In particular, interaction with the host substrate offers new ways to tune the properties of graphene and the case of functional transition metal oxides is of significant interest. For instance, graphene field-effect transistors (FETs) built on ferroelectric Pb(Zr,Ti)O$_3$ substrates display pronounced memory effects, ultra-low voltage operation~\cite{Gr-Ferro}, and open the way to novel nanoplasmonic devices~\cite{Gr-PZT-THz}. Similarly, graphene photo-sensitivity was shown to increase 25 times on TiO$_2$ substrates~\cite{Gr-TiO2} and intriguing magnetic phenomena are actively pursued in devices combining graphene with EuO substrates~\cite{Gr-EuO} or magnetic La$_x$Sr$_{1-x}$MnO$_3$ electrodes~\cite{Gr-LSMO}.

In the family of transition metal oxides, interfaces between the bulk band insulators LaAlO$_3$ (LAO) and SrTiO$_3$ (STO) occupy a special place due to their multiple, electric-field tunable properties, such as conductivity~\cite{oht-hwa, thiel}, superconductivity~\cite{Supcond-LAO, Au-LAO}, magnetism~\cite{magn-LAO, magn-storn} and spin-orbit coupling~\cite{soc-LAO}. These phenomena stem from the emergence of an interfacial two-dimensional electron system (2DES) when more than 3 unit cells (u.c.) of LAO are grown on a TiO$_2$-terminated STO crystal. Such 2DES is located only a few nm below the surface and its properties are therefore extremely sensitive to other materials deposited on this surface, such as metals~\cite{metal-LAO} or other oxides~\cite{oxide-LAO}. In this scenario, hybrid structures combining graphene with LAO/STO junctions represent an exciting platform in which novel phenomena may emerge from the strong electronic coupling of the respective 2DESs. In particular, collective interlayer-correlated phases driven by the strong Coulomb interactions are expected at low temperatures, in analogy to what observed in graphene/GaAs/AlGaAs systems~\cite{Gamucci}. Furthermore, magnetic or superconducting properties may be induced in graphene due to the proximity interaction with the ordered phases of the interfacial 2DES. In order to enable the exploration of these possibilities, the necessary starting point is a clear understanding of the transport behavior of these hybrid systems. LAO/STO-graphene systems, however, have been so far little studied. In a recent paper~\onlinecite{gr-LAO}, graphene was integrated with insulating LAO/STO substrates with subcritical LAO thickness, where a scanning probe microscope was used to ``write'' nanometer-wide conductive regions on the oxide system~\cite{AFM-litho} and graphene was used as a gate electrode controlling the conduction of these nanowires at the LAO/STO interface.

\begin{figure*}[t!]
	\begin{center}
		\includegraphics[width=0.75\textwidth]{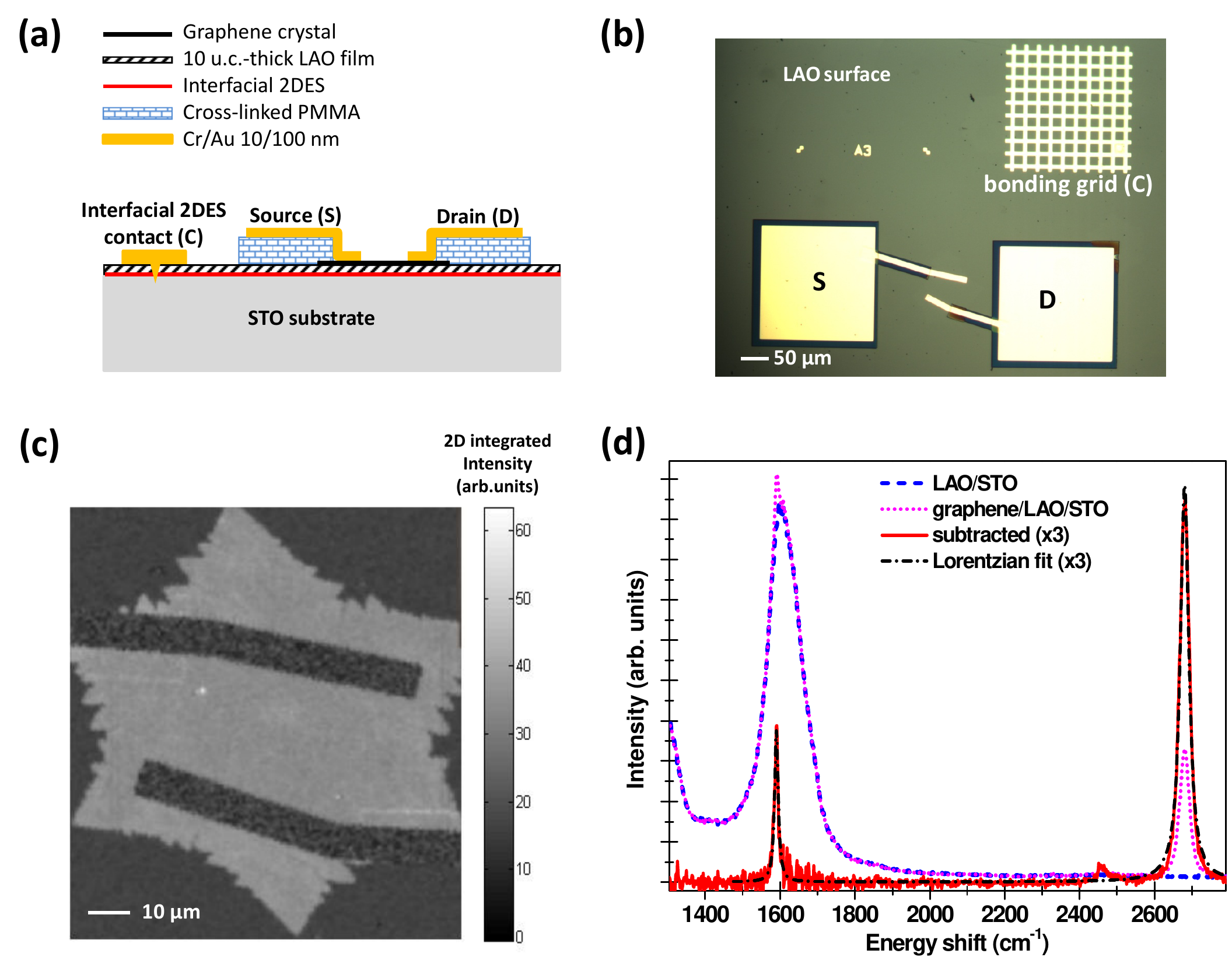}
		\caption{{\bf Monolayer of CVD graphene crystal deposited on LAO/STO.} (a) Sketch of the device architecture and (b) optical image of a representative three-terminal graphene/LAO/STO device for electrical transport studies at room temperature. Raman microspectrocopy of the contacted CVD graphene monocrystal in a fabricated device, showing the (c) map of the integrated intensity of the 2D peak and the (d) Raman spectroscopic analysis of the graphene crystal. The substrate-free graphene spectrum (red line) was obtained by subtracting Raman spectra acquired on regions covered by/free of graphene (dotted pink/dashed blue); the spectrum is compatible with a single layer crystal and the absence of D peak reveals that graphene's good quality is maintained during the fabrication procedure.}
		\label{device}
	\end{center}
\end{figure*}

Here we report on extended ($\approx 10^4\,{\rm \mu m^2}$) junctions between monolayer graphene grown by chemical vapor deposition (CVD) and a conductive, $4\,{\rm nm}$-deep LAO/STO interface hosting a 2DES. 
Our results demonstrate that strong electrostatic coupling with virtually no leakage can be obtained at room-temperature (RT) between a two-dimensional hole gas in graphene and the LAO/STO 2DES, as long as the graphene-oxide bias $V_{\rm GO}$ is sufficiently small ($|V_{\rm GO}|\lesssim1\,{\rm V}$). At larger values of $|V_{\rm GO}|$, we shall report strongly non-linear transport across the vertical graphene-LAO/STO junction that will be linked to tunneling currents between the two electron systems.

The device structure adopted in this work is illustrated in Fig.~1. High-quality 2DESs were obtained by growing 10 u.c.-thick LAO films on TiO$_2$-terminated STO chips by pulsed laser deposition~\cite{Miletto-growth}. Cr/Au alignment markers were fabricated by e-beam lithography and thermal evaporation and no additional processing was carried out in order to minimize contamination and damage of the oxide system. Single-crystalline graphene flakes typically measuring $100\,{\rm \mu m}$ in diameter were grown by CVD on flat, ex-situ passivated Cu foils~\cite{Vaidas}. Graphene crystals were finally deposited on the clean surfaces of LAO/STO heterostructures using a polymer-based transfer process~\cite{bubble} and localized thanks to the Cr/Au markers. The device architecture was designed taking particular care to avoid spurious electrical losses across the $\approx4\,{\rm nm}$-thin LAO barrier. An insulating layer of $200\,{\rm nm}$ of cross-linked PMMA was inserted between the LAO/STO and graphene electrodes (see the sketch of Fig.~\ref{device}a and the optical picture of Fig.~\ref{device}b), so that the only electrical path between the contact leads and the LAO/STO 2DES occurred vertically through the graphene-LAO/STO junction. The correct alignment of the flakes and their quality at the end of the process were assessed through Raman spectroscopy, as shown in Fig.~\ref{device}c,d for one of the studied devices. Given the relatively strong background signal from the LAO/STO substrate, Raman spectra were processed by subtracting the response of the bare oxide heterostructure (see Fig.~\ref{device}d). Once this background signal was removed, Raman emission clearly indicated that the graphene flake was a high-quality monolayer, with a single-Lorentzian 2D peak, the expected G/2D peak ratio for monolayer graphene and no discernible D peak throughout the crystal~\cite{Ram-Graph}. The correct alignment and uniformity of the flake is demonstrated in Fig.~\ref{device}c, where we report a Raman map integrated over the spectral region $2600-2800\,{\rm cm^{-1}}$ corresponding to the 2D peak.

\begin{figure}[ht!]
	\begin{center}
		\includegraphics[width=0.49\textwidth]{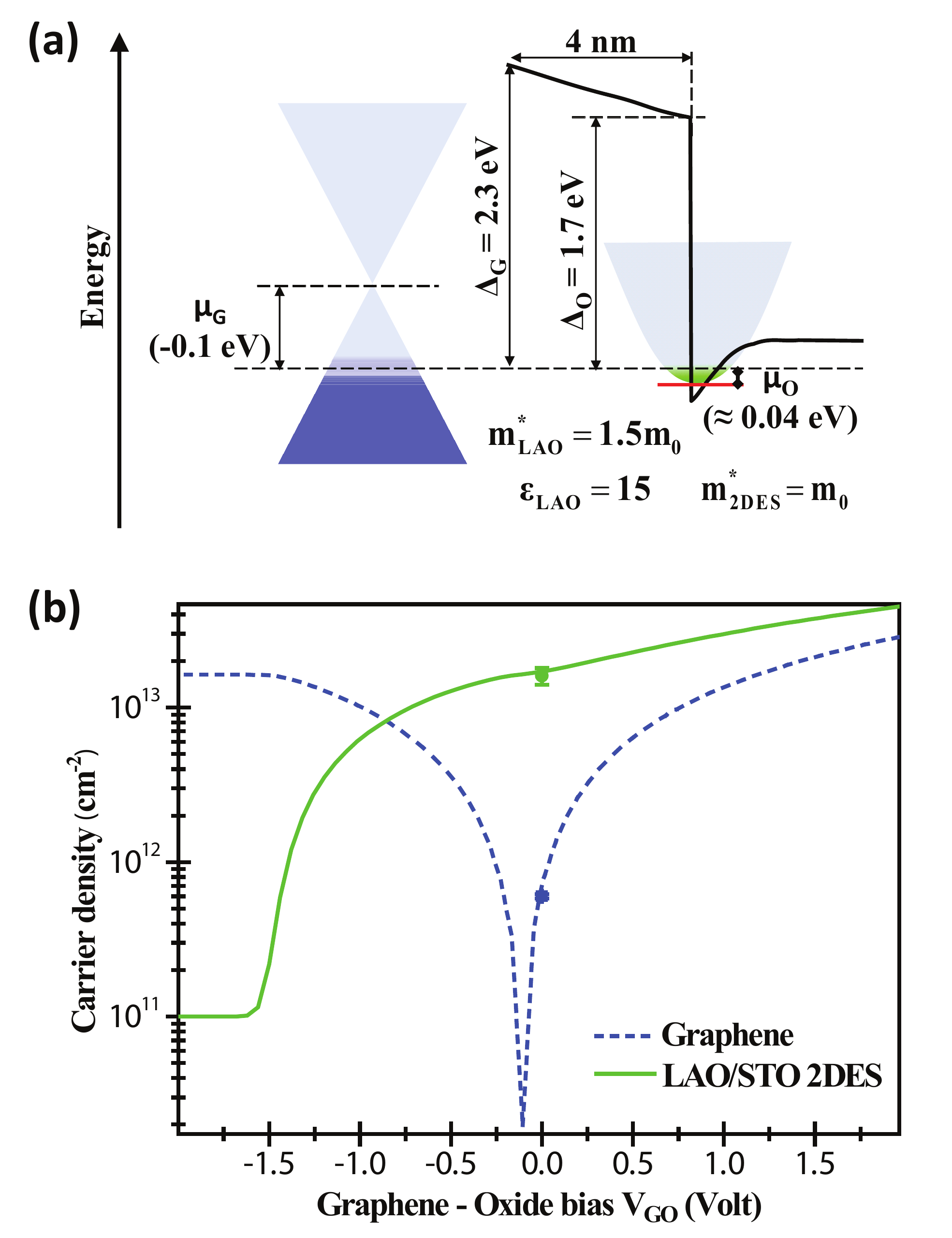}
		\caption{{\bf Theoretical modeling of graphene-LAO/STO systems.} (a) Schematic of the conduction band profile of the LAO/STO heterostructure and the relative band alignment between graphene and LAO barrier in equilibrium conditions, illustrating the relevant microscopic parameters of the vertical junction and the corresponding values that best reproduce the experimental data. (b) Evolution of the graphene (dashed blue) and interfacial 2DES (solid green) carrier density with vertical bias, showing the pronounced electric-field-effects in the hybrid structure. Dots represent the experimentally measured carrier densities for the two conductive systems in calibration multiterminal devices.}
		\label{teo-mod}
	\end{center}
\end{figure}

Device parameters and structure were chosen to achieve strong electrostatic coupling between the 2DESs in graphene and in the LAO/STO heterojunction, with negligible tunneling in the small interlayer bias ($V_{\rm GO}$) limit. Based on electron affinity values~\cite{EleAff}, we estimate a large ($\sim\,2\,{\rm eV}$) graphene~-~LAO barrier (see Fig.~\ref{teo-mod}a) that should electrically decouple the two electronic systems. At present many of the band parameters of the junction, in particular the exact band offsets and, consequently, the built-in field in the LAO layer, are still debated~[\onlinecite{offset},\onlinecite{treske}, and references therein]. In addition, the triangular quantum well potential at the LAO/STO interface confines the motion of oxide electrons to a non-trivial set of 2D subbands characterized by different and anisotropic effective masses~\cite{band-teo, LAO-X-ray}. In the following, the interface 2DES will be simply described by a single isotropic 2D subband and charge transport between the two 2DESs will be assumed to occur via direct tunneling across the $\approx4\,{\rm nm}$ LAO barrier. The electrostatic coupling between the two 2DES will be described in terms of a capacitor in which the electrodes have a finite density of states (DOS)~\cite{SuppNote}.

Figure~\ref{teo-mod}b reports the predicted evolution of the carrier density in the two 2DESs, setting the values at zero bias to the experimentally-measured ones~\cite{SuppNote}. For this measurement, multi-terminal Hall-bar structures were realized by a lift-off technique~\cite{lift-off} with a nominally identical 10 u.c. LAO/STO heterointerface. The electron densities at $4.2\,{\rm K}$ were found to be $1.6\,\times\,10^{13}\,{\rm cm^{-2}}$ and $-6\,\times\,10^{11}\,{\rm cm}^{-2}$ for the LAO/STO and the graphene 2DES respectively. The negative density indicates that the as-deposited graphene is p-doped, as typically observed. Based on this experimental input, the model predicts that graphene can be tuned up to densities of $\sim\,10^{13}\,{\rm cm}^{-2}$ both on the electron and hole sides and that LAO/STO can be fully depleted by biasing graphene at a few Volts. Such a strong electrostatic coupling is due to the small thickness of the LAO barrier: assuming $\kappa_{\rm LAO}\,=\,20$ for the LAO relative permittivity, the geometric capacitance per unit area is estimated to be $4.5\,{\rm \mu F/cm^2}$, corresponding to a gating efficiency $\approx400$ times higher than available in typical $300\,{\rm nm}$ thick oxidized Si. As a consequence, the charge-neutrality point (CNP) is expected to occur in graphene at very low (negative) bias values ($|V_{\rm GO}|<0.5\,{\rm V}$).

\begin{figure}[t!]
	\begin{center}
		\includegraphics[width=0.49\textwidth]{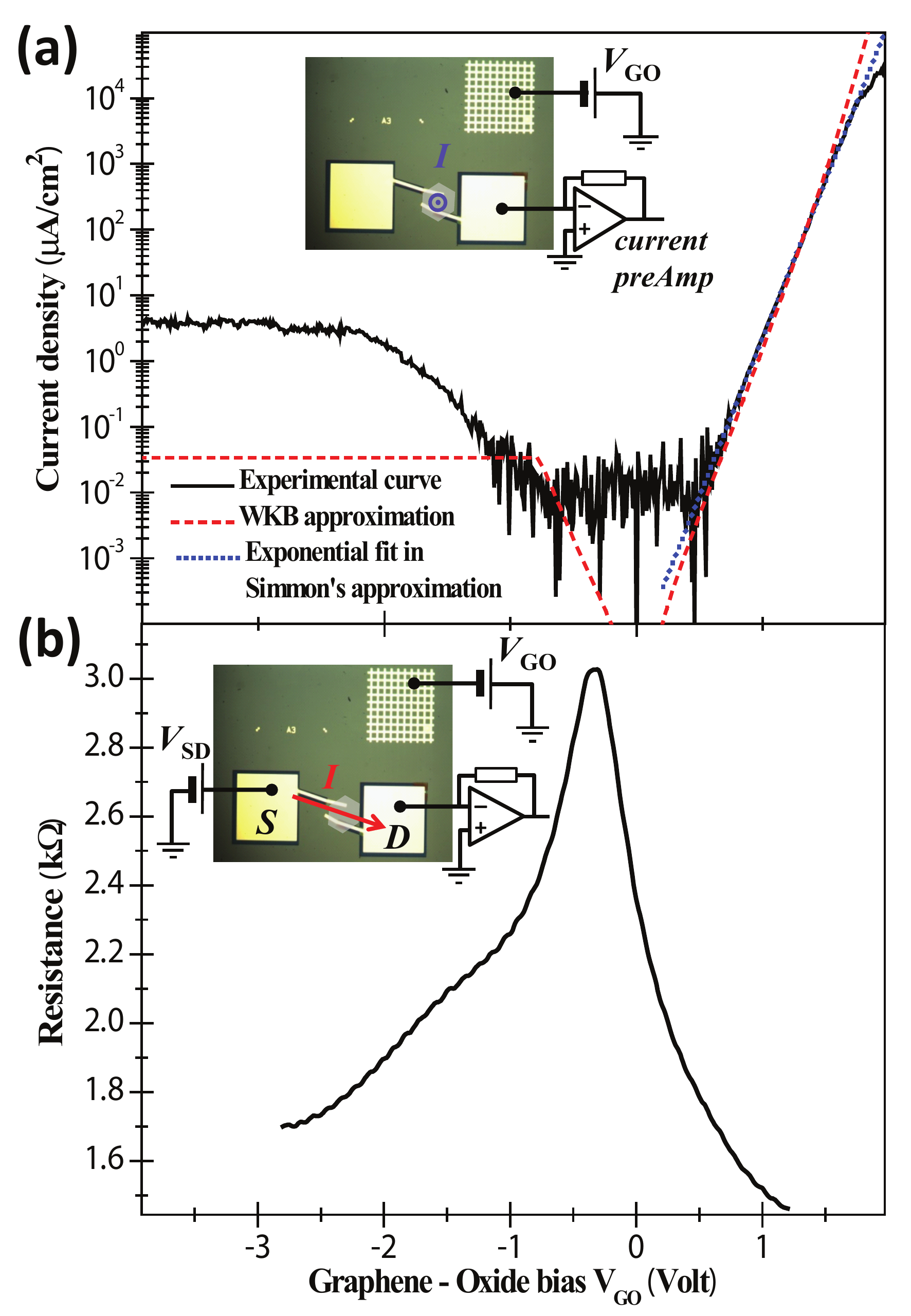}
		\caption{{\bf Pronounced non-linear transport and field-effects in graphene-LAO/STO systems.} (a) Graphene - oxide 2DES (black solid) current density as a function of vertical bias $V_{\rm GO}$ measured in a representative device at room temperature and the corresponding graphene-2DES tunneling characteristic calculated in the full WKB (red dashed) and Simmon's approximation (blue dotted)~\cite{SuppNote}. (b) Room temperature 2-point resistance of the single layer graphene crystal as a function of $V_{\rm GO}$, suggesting the tuning of graphene carrier density across the Dirac point. The insets are schematics of the electrical setups employed for studying the transport properties of the system. The blue and red arrows represent respectively the vertical and in-plane current flow in our device.} 
		\label{vertical}
	\end{center}
\end{figure}

The typical transport and gating properties of a representative graphene-LAO/STO hybrid device are reported in Fig.~\ref{vertical}. The insulating properties of the PMMA layer were preliminarily studied on $\approx 200\times 200\,{\rm \mu m^2}$ test pads covered by unconnected metallic electrodes: negligible ($\lesssim 0.1\,{\rm nA}$) vertical currents were observed for vertical biases in the studied $V_{\rm GO}$ range. Graphene devices were measured using a piezo-actuated microprobe system from Imina Technologies in order to avoid inducing electrical loss through the PMMA as a consequence of wire-bonding. We studied the RT vertical transport between the oxide interfacial 2DES and graphene by applying a DC bias $V_{\rm GO}$ to the graphene electrode while holding the interfacial 2DES at ground. As shown in Fig.~\ref{vertical}a, a strongly non-linear transport characteristic was observed, where different regimes can be distinguished. Around zero bias, graphene and the 2DES are insulated and no current could be detected above the noise: a current below few pA is measured for $|V_{\rm GO}|\,<\,0.7\,{\rm V}$, corresponding to a resistance in the range $100\,{\rm G\Omega}-1\,{\rm T\Omega}$. Based on the measured carrier densities, graphene's CNP is expected to lie in this voltage range (see Fig.~\ref{teo-mod}b). Outside this interval, vertical transport sets on, with an approximately exponential current-voltage characteristic. For $V_{\rm GO}\,<\,-2.5\,{\rm V}$, vertical current saturates to a fraction of nA: as discussed in the following, this is due to a field-effect depletion of the oxide interfacial 2DES. The exponential growth is much more pronounced for positive bias values: the current grows by 5 orders of magnitude in less than 1V of bias change and it is ultimately limited (for $V_{\rm GO} > 2\,{\rm V}$) by the resistive load of the 2DES paths connecting the junction area to the electrodes.

Transport in the latter regime is governed only by the vertical junction properties and was therefore amenable to theoretical modeling~\cite{SuppNote}. The observed transport characteristics at positive bias are compatible with direct tunneling of electrons between graphene and the interfacial 2DES. In fact, the large graphene-LAO barrier ($\sim\,2\,{\rm eV}$) suggests that thermionic emission should be suppressed even at RT, while tunneling remains possible because of the ultra-thin barrier. Zener tunneling across the LAO should be negligibly small for 10 u.c.-thick LAO films~\cite{tunn-Pt-LAO}, particularly for positive values of $V_{\rm GO}$, which tend to flatten the LAO potential. Therefore, only direct graphene-2DES tunneling was considered and calculated in the semiclassical (WKB) approximation, that was already successfully adopted to describe transport features in both metal/LAO/STO~\cite{tunn-Pt-LAO} and van der Waals heterostructures comprising graphene~\cite{teo-Gr-tunn}. Input parameters used to reproduce the experimental data are reported in Fig.~\ref{teo-mod}a and were set at the typically reported values in the literature, apart from the barrier effective mass that was obtained by fitting the data in the Simmon's approximation (dotted blue curve in Fig.~\ref{vertical}a)~\cite{SuppNote}. Fig.~\ref{vertical}a compares the resulting theoretical prediction (red dashed line) with a typical experimental curve (solid black line): the model captures the main transport features in the whole bias range and it is quantitatively accurate in the intrinsic regime, thus providing evidence that graphene~-~interfacial 2DES transport is indeed dominated by direct tunneling of electrons through the LAO barrier. The remaining discrepancies between the theoretical prediction and experimental data in the intrinsic regime point to the possible role played by the inner structure of the oxide interfacial 2DES~\cite{band-teo, LAO-X-ray} (the population of several subbands changes the transport characteristics), or the possible contribution of tunneling processes assisted by phonons and/or in-gap defect states in the LAO barrier (e.g.~O vacancies). The deviations observed for $V_{\rm GO}\,<\,0.5\,{\rm V}$ are due to electronic noise and interface depletion issues and are not addressed by the present model. Finally, we note that similar rectifying behavior was already reported in metal(Pt,Au)/LAO/STO junctions~\cite{tunn-Pt-LAO, Au-LAO} and was also attributed mainly to quantum tunneling~\cite{tunn-Pt-LAO}. The similarities are due to the metallic behavior of graphene for large positive values of $V_{\rm GO}$, as shown in Fig.~\ref{teo-mod}b, and the comparable electron affinities of graphene and Pt, Au.

In the low inter-layer bias limit, and for the full range $V_{\rm GO}\,<\,1.0\,{\rm V}$, graphene and the interfacial 2DES can be considered electrically insulated, with vertical currents lower than $0.5\,{\rm nA}$. The existence of this quasi-insulating regime makes it possible to explore Coulomb coupling between the two electron systems. We monitored the 2-point RT resistance of graphene as a function of $V_{\rm GO}$: the result is shown in Fig.~\ref{vertical}b for a representative device and demonstrates the significant modulation of the graphene resistivity by the electric field of the 2DES electrode. The resistance curve indicates that graphene can be easily tuned between p-type and n-type conduction at a $V_{\rm GO}$ value as little as a fraction of a Volt, despite the non-negligible intrinsic doping $-6\times 10^{11}\,{\rm cm^{-2}}$. This behavior confirms the extremely large gating efficiency of the system.

The analogous field-effect of graphene on the interfacial 2DES predicted from the capacitor model in Fig.~\ref{teo-mod}b can be inferred both from gating and vertical transport features. In fact, the graphene gating efficiency experiences a sudden drop for ${\rm V_{\rm GO}}\,<\,-1{\rm V}$ and is totally suppressed for $V_{\rm GO}\,<\,-2.5{\rm V}$, as signaled by the resistivity saturation in Fig.~\ref{vertical}b. This effect matches the kink and the subsequent current saturation in the vertical transport curve in Fig.~\ref{vertical}a, which persists up to $V_{\rm GO}\,=\,-90\,{\rm V}$ (data not shown). These observations provide experimental evidence of the predicted depletion of the 2DES for large negative bias; the creation of an insulating region below graphene interrupts the conducting paths to the LAO/STO electrodes and therefore limits vertical transport.

In conclusion, we reported on the transport properties of the graphene-LAO/STO hybrid system and showed that it is characterized by strong electrostatic coupling at low inter-layer bias and direct tunneling coupling at large bias values. Efficient gating could be fruitfully exploited for the realization of hybrid ``dual'' FETs~\cite{dual-FET} with very-low operating voltage, where graphene can be employed either as gate, or as the conducting channel back-gated by the 2DES. The present device architecture can be relevant to the investigation of collective phases with very strong inter-layer correlations at low temperatures.

The research leading to these results received funding from the European Union Seventh Framework Programme under grant agreement No. 604391 Graphene Flagship. S.R. acknowledges the support of CNR through the bilateral CNR-RFBR project 2015-2017.

\balancecolsandclearpage

	\setcounter{equation}{0}
	\setcounter{figure}{0}
	\setcounter{section}{0}
	\setcounter{page}{1}
	\makeatletter
	\renewcommand{\theequation}{SE\arabic{equation}}
	\renewcommand{\thepage}{Sp\arabic{page}}
	\renewcommand{\thesection}{\Roman{section}}
	\renewcommand{\bibnumfmt}[1]{[S#1]}
	\renewcommand{\citenumfont}[1]{S#1}

\section{Electrostatic considerations}
The vertical voltage bias $V_{\rm GO}$ is defined as the electro-chemical potential difference  between the two-dimensional electron system (2DES) in graphene and the 2DES at the oxide interface (the ``interfacial 2DES''), i.e.
\begin{equation}\label{eq:voltage}
V_{\rm GO} \equiv \phi_{\rm G} + \frac{\mu_{\rm G}+E^{(0)}_{\rm G}}{-e} - \phi_{\rm O} - \frac{\mu_{\rm O}+E^{(0)}_{\rm O}}{-e}~,
\end{equation}
where $-e$ is the electron charge, $\phi_{{\rm G}({\rm  O})}$ is the electric potential of the graphene sheet (interfacial 2DES), $E^{(0)}_{{\rm G}({\rm  O})}$ is the energy of the Dirac point (bottom of the 2D subband, represented by the red line inside the triangular well in Fig.~2a) of graphene (the interfacial 2DES) measured from the vacuum level, and $\mu_{{\rm G}({\rm O})}$ is the corresponding chemical potential measured with respect to the Dirac point (2D subband edge).

The two 2DESs can be modeled as a parallel plate capacitor. The electric potential difference can therefore be written as:
\begin{equation}
\phi_{\rm G} - \phi_{\rm O}  = \frac{ed(n_{\rm G} - n_{\rm O})}{2 \epsilon_{0}\epsilon_{\rm r}}+\tilde{\phi}\equiv \frac{e(n_{\rm G} - n_{\rm O})}{2 C}+\tilde{\phi}~.
\end{equation}
Here, $n_{{\rm G}({\rm O})}$ is the density of {\it free} electrons in the graphene sheet (interfacial 2DES), $\epsilon_{\rm r}$ is the relative dielectric constant of the barrier material, $\tilde{\phi}$ is a contribution arising from bound charges in the system (e.g. charges bounded at the termination surface of LAO, charged impurities adsorbed on the graphene flake, etc.), and $C=\epsilon_0\epsilon_{\rm r}/d$ is the geometrical capacitance of the junction per unit area.

In the $T\to 0$ limit the chemical potential coincides with the Fermi energy,
\begin{eqnarray}
\mu_{\rm G}(n_{\rm G},T=0)=\hbar v_{\rm F}\sqrt{\pi n_{\rm G}}\label{eq:chempotgraphene}\\
\mu_{\rm O}(n_{\rm O},T=0)=\frac{\pi \hbar^2}{m_{\rm b}}n_{\rm O}\label{eq:chempot2DES}~.
\end{eqnarray}
In writing the previous equation we have dramatically simplified the description of the interfacial 2DES: we assume that it occupies a single 2D subband of the interfacial triangular quantum well, and assume the subband dispersion to be parabolic and isotropic with an effective band mass $m_{\rm b}$. Following Ref.~\onlinecite{Tolsma2015}, we identify the band mass with the mass of light quasiparticles and set $m_{\rm b}\sim m_{\rm e}$, where $m_{\rm e}$ is the bare electron mass in vacuum. In Eq.~(\ref{eq:chempotgraphene}), $v_{\rm F}\sim 10^6~{\rm m}/{\rm s}$ is the graphene Fermi velocity. At finite temperature $T$, $\mu_{\rm O}(n_{\rm O},T)$ and $\mu_{\rm G}(n_{\rm G},T)$ can be found numerically. 
Increasing temperature, the chemical potential moves towards (away from) the Dirac point (subband edge) in graphene (the interfacial 2DES).

At zero bias, Eq.~(\ref{eq:voltage}) reduces to
\begin{equation}
\begin{split}
&0=\frac{e(n^{(0)}_{\rm O} - n^{(0)}_{\rm G})}{2 C}-\tilde{\phi}+\frac{\mu_{\rm O}^{(0)}
	+E^{(0)}_{\rm O}-\mu_{\rm G}^{(0)}-E^{(0)}_{\rm G}}{e}~,
\end{split}
\end{equation}
where $n^{(0)}_{{\rm O}({\rm G})}$ are the electron densities at zero bias, while $\mu^{(0)}_{{\rm O}({\rm G})}=\mu_{{\rm O}({\rm G})}(n^{(0)}_{{\rm O}({\rm G})},T)$ are the chemical potentials at zero bias.

We now assume that, increasing the bias voltage, charge is transferred only between graphene and the interfacial 2DES. 
We write the carrier densities as $n_{{\rm G}({\rm O})}=n^{(0)}_{{\rm G}({\rm O})} \mp \delta n$, where the minus (plus) sign applies to graphene (the interfacial 2DES). We can therefore write an equation for the unknown $\delta n$, which can be easily solved numerically. It reads as following
\begin{equation}
\begin{split}
V_{\rm GO}=&\frac{e\delta n}{C}+\frac{\mu_{\rm G}(n^{(0)}_{\rm G} - \delta n,T)-\mu_{\rm G}^{(0)}}{-e}\\
&-\frac{\mu_{\rm O}(n^{(0)}_{\rm O} + \delta n,T) - \mu_{\rm O}^{(0)}}{-e}~.
\end{split}
\end{equation}
The input parameters that are needed to solve this equation are: i) the capacitance per unit area $C$, ii) the electron densities $n^{(0)}_{{\rm G}({\rm O})}$ at zero bias, iii) the effective band mass $m_{\rm b}$ of the interfacial 2DES, and iv) temperature $T$.

The solution of this equation for $n^{(0)}_{\rm G}=-0.7\times10^{11}{\rm cm^{-2}}$, $n^{(0)}_{\rm O}=1.6 \times 10^{13}{\rm cm^{-2}}$, $C=3.3~{\rm \mu F}/{\rm cm}^{2}$, and $T=300~{\rm K}$ is reported in Fig.~2b of the main text.

\section{Theory of the inter-layer tunneling current between graphene and the interfacial 2DES}
The current density flowing from graphene to the interfacial 2DES can be written as:
\begin{equation}\label{eq:tunneling_current}
J=\frac{e}{S}\sum_{\alpha \in {\rm O}}\sum_{\beta \in {\rm G}}W_{\alpha\beta}(n_\alpha-n_\beta)~,
\end{equation}
where $\alpha$ ($\beta$) labels the electronic states localized at the oxide interface (in the graphene layer), $n_{\alpha(\beta)}$ is the occupation probability of the state $\alpha$ ($\beta$), $W_{\alpha\beta}$ is the transition probability per unit time from the state $\alpha$ to the state $\beta$, and $S$ is the surface of the device. 

We assume that: i) the occupation probability of the states is given, in each electron system, by a Fermi-Dirac distribution $n_{\alpha(\beta)}=f_{\rm FD}(\epsilon_{\alpha(\beta)}-\mu_{{\rm O}({\rm G})})$ where 
$f_{\rm FD}(E) = [\exp{(
	E/(k_{\rm B}T))} +1]^{-1}$, and $\epsilon_{\alpha(\beta)}$ is measured from the bottom of the conduction band (from the Dirac point), $T$ being temperature;
ii) the transition probability $W_{\alpha \beta}$ depends only on the energy of the initial and final states; iii) only elastic processes contribute to the tunneling current. 
Under these assumptions $W_{\alpha \beta}=\delta(\epsilon_\alpha+E_{\rm O}^{(0)}-e\phi_{\rm O}-\epsilon_\beta-E_{\rm G}^{(0)}+e\phi_{\rm G})w(\epsilon_\alpha)/S$ and  Eq.~(\ref{eq:tunneling_current}) can be simplified to:
\begin{widetext}
	\begin{equation}\label{eq:tunneling_current_processed}
	J =\frac{e}{S^2}\sum_{\alpha \in {\rm O}}\sum_{\beta \in {\rm G}}\int_{-\infty}^{\infty} d\epsilon~\delta\left(\epsilon-\epsilon_\alpha-\frac{e^2 \delta n}{2C}+\mu_{\rm O}^{(0)}\right)\delta\left(\epsilon-\epsilon_\beta+\frac{e^2 \delta n}{2C}+\mu_{\rm G}^{(0)}\right)w(\epsilon_\alpha)[f_{\rm FD}(\epsilon_{\alpha}-\mu_{\rm O}) - 
	f_{\rm FD}(\epsilon_{\beta}-\mu_{\rm G})]~.
	\end{equation}
\end{widetext}
In deriving Eq.~(\ref{eq:tunneling_current_processed}) we have used the following identity
\begin{equation}
\begin{split}
&\delta(\epsilon_\alpha+E_{\rm O}^{(0)}-e\phi_{\rm O}-\epsilon_\beta-E_{\rm G}^{(0)}+e\phi_{\rm G})=\\
&\int_{-\infty}^{\infty} d\epsilon\delta(\epsilon-\epsilon_\alpha-\frac{e^2 \delta n}{2C}+\mu_{\rm O}^{(0)})\delta(\epsilon-\epsilon_\beta+\frac{e^2 \delta n}{2C}+\mu_{\rm G}^{(0)})~.
\end{split}
\end{equation}
We approximate the transmission probability with that of a trapezoidal potential barrier:
\begin{equation}
w(\epsilon_\alpha)=\frac{2 \pi}{\hbar}\Gamma^2T_{\rm TB}(\epsilon_\alpha,\Delta_O+\mu_{\rm O}^{(0)},\Delta_O+\mu_{\rm O}^{(0)}-e(\phi_{\rm G}-\phi_{\rm O}))~,
\end{equation}
where $\Gamma$ quantifies the probability of jumping from one of the two electron systems to the barrier, $\Delta_{\rm O}$ is the height of the barrier at the oxide interface measured from the Fermi level at zero bias, $T_{\rm TB}$ is the transmission probability through a trapezoidal barrier calculated using the Wentzel-Kramers-Brillouin (WKB) approximation:
\begin{equation}
\begin{split}
&T_{\rm TB}(\epsilon,\Delta_1,\Delta_2)=\\
&=\exp \left[-\frac{\alpha}{d} \int_0^{x^*}dx\sqrt{\frac{d\Delta_1+(\Delta_2-\Delta_1)x}{d}-\epsilon}\right]\\
&=\exp\left[-\frac{2 \alpha}{3}\Theta(\Delta_2-\epsilon)\frac{(\Delta_2-\epsilon)^{3/2}}{\Delta_2-\Delta_1}\right]\\
&\times \exp\left[-\frac{2 \alpha}{3}\Theta(\Delta_1-\epsilon)\frac{(\Delta_1-\epsilon)^{3/2}}{\Delta_1-\Delta_2}\right]~.
\end{split}
\end{equation}
where $\Theta(x)$ is the usual Heaviside step function and $\alpha= 2\sqrt{2 m^*}d/\hbar$ with $ m^*$ being the conduction band effective mass in the LAO barrier, $x^*$ is the classical turning point of the electron trajectory, while $\Delta_{1(2)}$ are the heights of the two sides of the barrier.

Introducing the density-of-states per unit area of the interfacial 2DES and graphene $D_{{\rm O}({\rm G})}(\epsilon)=S^{-1}\sum_{\alpha\in {\rm O}({\rm G})} \delta(\epsilon-\epsilon_\alpha)$ or, more explicitly, $D_{\rm O}(\epsilon)=m_{\rm b}/(\pi \hbar^2)$, and $D_{\rm G}(\epsilon)=2|\epsilon|/(\pi \hbar^2 v_{\rm F}^2)$, we find the final result for the tunneling current:
\begin{widetext}
	\begin{equation}\label{eq:tunneling_current_final}
	\begin{split}
	J=\frac{2 \pi e \Gamma^2}{\hbar}\int_{-\infty}^{\infty}d\epsilon~&D_{\rm O}\left(\epsilon+\mu_{\rm O}^{(0)}-\frac{e^2 \delta n}{2C}\right)
	D_{\rm G}\left(\epsilon+\mu_{\rm G}^{(0)} + 
	\frac{e^2 \delta n}{2C}\right)
	T_{\rm TB}\left(\epsilon,\Delta_{\rm O}+\frac{e^2 \delta n}{2C},\Delta_{\rm G}-\frac{e^2 \delta n}{2C}\right)
	\\
	&\times\left[f_{\rm FD}\left(\epsilon-\mu_{\rm O} + \mu_{\rm O}^{(0)}-\frac{e^2 \delta n}{2C}\right)-f_{\rm FD}\left(\epsilon-\mu_{\rm G}+ \mu_{\rm G}^{(0)}+\frac{e^2 \delta n}{2C}\right)\right]~.
	\end{split}
	\end{equation}
\end{widetext}
In writing the previous equation, we have introduced the barrier height at the graphene-LAO interface with respect to the common Fermi level at zero bias $\Delta_{\rm G}=\Delta_{\rm O}+\tilde{\phi}-e^2(n_{\rm O}^{(0)}-n_{\rm G}^{(0)})/(2C)$. 
Eq.~(\ref{eq:tunneling_current_final}) represents the main result of our model and has been used to generate the vertical transport curve shown in red dashed line in the Fig.~3a of the main text.

We now make some simplifying assumptions to obtain an approximate analytical result for the tunneling current density $J$, in order to both compare it with well-known formulas for 3D-to-3D tunneling~\cite{Simmons1963a,Simmons1963b,Holm1951,Fowler1928} and use it to fit the experimental data.

Neglecting quantum capacitance effects (i.e.~the variation of the chemical potentials $\mu_{{\rm G}({\rm O})}$ with density) Eq.~(\ref{eq:tunneling_current_final}) reduces to
\begin{widetext}
	\begin{equation}
	\begin{split}
	J=\frac{2 \pi e \Gamma^2}{\hbar}
	\int_{-\infty}^{\infty}d\epsilon~&
	D_{\rm O}\left(\epsilon+\mu_{\rm O}^{(0)}-\frac{eV_{\rm GO}}{2}\right)
	D_{\rm G}\left(\epsilon+\mu_{\rm G}^{(0)}+\frac{eV_{\rm GO}}{2}\right)
	T_{\rm TB}\left(\epsilon,\Delta_{\rm O}+\frac{eV_{\rm GO}}{2},\Delta_{\rm G}-\frac{eV_{\rm GO}}{2}\right)
	\\
	&
	\times\left[
	f_{\rm FD}\left(\epsilon-\frac{eV_{\rm GO}}{2}\right)
	-f_{\rm FD}\left(\epsilon+\frac{eV_{\rm GO}}{2}\right)
	\right]~.
	\end{split}
	\end{equation}
\end{widetext}
Setting $T=0$ and approximating the density-of-states of each 2DES with its value calculated at the chemical potential, we obtain
\begin{equation}\label{eq:TunnelingCurrentApprox}
J\approx A \int_{-eV_{\rm GO}/2}^{eV_{\rm GO}/2}d\epsilon~T_{\rm TB}\left(\epsilon,\Delta_{\rm O}+\frac{eV_{\rm GO}}{2},\Delta_{\rm G}-\frac{eV_{\rm GO}}{2}\right)~,
\end{equation}
where $A\equiv 2\pi e \Gamma^2  D_{\rm O}(\mu_{\rm O}^{(0)})
D_{\rm G}(\mu_{\rm G}^{(0)})/\hbar$.

For low bias (i.e.~$|V_{\rm GO}|\ll \Delta_{\rm O}, \Delta_{\rm G}$), in the spirit of Simmons' formula~\cite{Simmons1963a,Simmons1963b,Holm1951}, we can replace the trapezoidal barrier with a rectangular barrier with average height $\bar{\Delta}=(\Delta_{\rm O}+\Delta_{\rm G})/2$ and transmission probability
\begin{equation}
T_{\rm RB}(\epsilon,\bar{\Delta})=\exp\left(-\alpha\Theta(\bar{\Delta}-\epsilon)\sqrt{\bar{\Delta}-\epsilon}~\right)~.
\end{equation} 
We find
\begin{widetext}
	\begin{equation}\label{eq:Simmons}
	J\approx A
	\frac{2}{\alpha^2}\left[\left(1+\alpha\sqrt{\bar{\Delta} -\frac{eV_{\rm GO}}{2}}\right) e^{-\alpha\sqrt{\bar{\Delta} -\frac{eV_{\rm GO}}{2}}}
	-\left(1+\alpha\sqrt{\bar{\Delta} +\frac{eV_{\rm GO}}{2}}\right) e^{-\alpha\sqrt{\bar{\Delta} +\frac{eV_{\rm GO}}{2}}}\right]~,
	\end{equation}
\end{widetext}
which generalizes Simmons' formula to the problem of tunneling between two 2DESs with no in-plane momentum conservation.

For large negative bias (i.e.~$eV_{\rm GO}\ll-\Delta_{\rm G}$), we can instead approximate Eq.~(\ref{eq:TunnelingCurrentApprox}) neglecting the smallest of the two barrier heights, i.e.~setting $\Delta_{\rm O} \to 0$. We find
\begin{equation}\label{eq:Fowler}
\begin{split}
J&\approx A \int_{-eV_{\rm GO}/2}^{eV_{\rm GO}/2}d\epsilon~\exp\left[-\frac{2 \alpha}{3}\frac{(\Delta_{\rm G} - eV_{\rm GO}/2-\epsilon)^{3/2}}{\Delta_{\rm G}-eV_{\rm GO}}\right]\\
&\approx A\frac{eV_{\rm GO}}{2}\exp\left[\frac{2\alpha}{3}\frac{\Delta_{\rm G}^{3/2}}{eV_{\rm GO}}\right]~.
\end{split}
\end{equation}
This expression is the 2D-to-2D equivalent of the Fowler-Nordheim equation~\cite{Fowler1928}.
However, in our case, the Fowler-Nordheim regime is preempted by the depletion of the interfacial 2DES.

\section{Analytical fitting procedure for positive bias}
\label{sect:fitting}
Eq.~(\ref{eq:tunneling_current_final}) reveals a very complicated dependence of the tunneling current on applied bias, which hinders the reproduction of the experimental data through direct fitting; therefore an indirect route was chosen to reproduce the transport curves, as described in the following for the representative curve in Fig.~3a in the main text.

We notice that between the noise floor and the onset of resistive saturation, i.e. for ${\rm 0.7} < V_{\rm GO} < 1.8~{\rm V}$ for the particular curve shown in Fig.~3a, the $J\,-\,V_{\rm GO}$ relationship reduces to a simple exponential function, as indicated by the linear dependence exhibited in the semi-logarithmic plot. The blue dotted curve in Fig.~3a represents the best fit to the data in this bias range with a two-parameter exponential function $J=c_1\exp (c_2 V_{\rm GO})$, yielding $c_1=3.6\times 10^{-5} {\rm \mu A/cm^2}$ and $c_2=11~{\rm V^{-1}}$.

To compare this result with our theoretical model, we need a simplified version of Eq.~\ref{eq:Simmons}. To this end, we notice that for typical values of $\alpha \approx 40$ and positive bias only the first exponential term in Eq.~\ref{eq:Simmons} is relevant, the second being suppressed by several orders of magnitude.
Fixing a reference point $V^*=1.3~{\rm V}$ at the center of the relevant bias interval and expanding to first order in $V_{\rm GO}-V^*$, we obtain
\begin{equation}\label{eq:exponential}
J\approx {\rm Const}(A,\alpha,\bar{\Delta},V^*) \times \exp\left(\frac{\alpha e V_{GO}}{4 \sqrt{\bar{\Delta} -\frac{eV^*}{2}}}\right)~,
\end{equation}
so that we can identify $c_2$ with $\alpha e/\left(4 \sqrt{\bar{\Delta} -eV^*/2}\right)$. We estimate an average barrier height $\bar{\Delta}\approx2~{\rm eV}$ and consequently extract a value of $m^*=1.5 m_{\rm e}$, which is significantly larger than previously reported values~\cite{Supp-tunn-Pt-LAO, mLAO}. The barrier is estimated in the following way: combining the reported electron affinities values of graphene ($4.6\,{\rm eV}$)~\cite{Supp-Gr-ele-aff} and LAO ($2.4\,{\rm eV}$)~\cite{Supp-tunn-Pt-LAO, Supp-LAO-ele-aff} and the experimentally measured $n^{(0)}_{\rm G}$, yields $\Delta_{\rm G}=2.3~{\rm eV}$; furthermore, by assuming typical values for the conduction band offset ($2\,{\rm eV}$) at the oxide interface~\cite{offset}, the triangular well confinement energy ($0.1\,-\,0.2\,{\rm eV}$), the interfacial subband effective mass $m_{\rm b}=m_{\rm e}$~\cite{Tolsma2015,Supp-tunn-Pt-LAO} and using the experimentally measured $n^{(0)}_{\rm O}$, we obtain $\Delta_{\rm O}\,\approx\,1.8~{\rm eV}$. The parameters concerning the electrostatics of the junction, i.e $d=4~{\rm nm}$, $\epsilon_{\rm r} = 15$, and the zero bias densities of the two electron systems were chosen the same as in Fig.~2. With these parameters, we generated the RT vertical transport characteristics in the full WKB approximation shown in dashed red line in Fig.~3a. The curve reproduces qualitatively the observed transport behavior in the whole bias range and is quantitatively accurate in the instrinsic bias range, providing evidence that the vertical graphene-oxide transport is dominated by direct tunneling of electrons through the LAO barrier.

\end{document}